\documentclass[twocolumn, switch]{article} 
\usepackage[colorlinks = true,
            linkcolor = purple,
            urlcolor  = blue,
            citecolor = cyan,
            anchorcolor = black]{hyperref}	
\usepackage{preprint}

\usepackage{amsmath, amsthm, amssymb, amsfonts}

\usepackage[numbers,square]{natbib}
\usepackage[utf8]{inputenc}	
\usepackage[T1]{fontenc}	
\usepackage{xcolor}		
\usepackage{booktabs} 		
\usepackage{nicefrac}		
\usepackage{microtype}		
\usepackage{lineno}		
\usepackage{float}			

\usepackage{lipsum}		

\usepackage{newfloat}
\DeclareFloatingEnvironment[name={Supplementary Figure}]{suppfigure}
\usepackage{sidecap}
\sidecaptionvpos{figure}{c}

\usepackage{titlesec}
\titlespacing\section{0pt}{12pt plus 3pt minus 3pt}{1pt plus 1pt minus 1pt}
\titlespacing\subsection{0pt}{10pt plus 3pt minus 3pt}{1pt plus 1pt minus 1pt}
\titlespacing\subsubsection{0pt}{8pt plus 3pt minus 3pt}{1pt plus 1pt minus 1pt}

\title{Development of planar microstrip resonators for electron spin resonance spectroscopy}

\usepackage{xwatermark}
\newwatermark[firstpage,color=gray!90,angle=0,scale=0.28, xpos=0in,ypos=-5in]{*correspondence: \texttt{chiranjib@iiserkol.ac.in}}

\usepackage{authblk}

\author[1, a]{Subhadip Roy}
\author[1, a]{Sagnik Saha}
\author[1]{Jit Sarkar}
\author[1]{Chiranjib Mitra}
\affil[1]{Department of Physical Sciences, Indian Institute of Science Education And Research Kolkata,India.}

\begin{document}

\twocolumn[ 
  \begin{@twocolumnfalse} 
  
\maketitle

\begin{abstract}
This work focuses on the development of planar microwave resonators which are to be used in electron spin resonance spectroscopic studies. Two half wavelength microstrip resonators of different geometrical shapes, namely straight ribbon and omega, are fabricated on commercially available copper clad microwave laminates.  Both resonators have a characteristic impedance of 50 $\Omega$.We have performed electromagnetic field simulations for the two microstrip resonators and have extracted practical design parameters which were used for fabrication. The effect of the geometry of the resonators on the quasi-transverse electromagnetic (quasi-TEM) modes of the resonators is noted from simulation results. The fabrication is done using optical lithography technique in which laser printed photomasks are used. This rapid prototyping technique allows us to fabricate resonators in a few hours with accuracy up to 6 mils. The resonators are characterized using a Vector Network Analyzer. The fabricated resonators are used to standardize a home built low-temperature continuous wave electron spin resonance (CW-ESR) spectrometer which operates in S-band, by capturing the absorption spectrum of the free radical DPPH, at both room temperature and 77 K. The measured value of g-factor using our resonators is consistent with the values reported in literature. The designed half wavelength planar resonators will be eventually used in setting up a pulsed electron spin resonance spectrometer by suitably modifying the CW-ESR spectrometer.
\end{abstract}
\vspace{0.35cm}

  \end{@twocolumnfalse} 
] 



\section{Introduction}
\label{intro}
Electron spin resonance (ESR) spectroscopy is a versatile technique which can be used to characterize the energy levels of an electron spin system and to obtain information about the dynamics of the system. The continuous wave electron spin resonance (CW ESR) spectroscopy variant is used for material characterization purposes \cite{CWESR}, whereas, the pulsed electron spin resonance spectroscopy is used for measuring spin-lattice and spin-spin relaxation times of the spin system \cite{RefB1}. In recent times pulsed ESR spectroscopy has drawn significant attention from the perspective of quantum information processing and can be used to perform various quantum information protocols in a spin system \cite{pulsed1}-\cite{pulsed2}. Microwave resonators form a vital part of the instrumentation for both variants of ESR spectrometers.\\
In general, a three-dimensional cavity resonator is employed in an electron spin resonance spectrometer to hold the sample and to concentrate the microwave magnetic field at the sample position. This variety of microwave resonators have high quality (Q) factor which in turn leads to high sensitivity of an ESR spectrometer \cite{RefB2}-\cite{RefJ1}. However, from the aspect of instrumentation of a pulsed ESR, the resonators should have low Q factor to ensure a fast response to the applied microwave pulses \cite{lowq}-\cite{RefJ2}. Planar microstrip resonators fit the above requirement as they can be easily designed to have low Q factors. The resonators also provide a region of concentrated uniform magnetic field where the sample under study can be placed easily in small quantity. An added advantage is that they can be fabricated easily with the existing printed circuit board manufacturing technologies \cite{Ghirri}.\\ 
For our case we have made the planar resonators after running detailed electromagnetic simulations. In subsequent sections we will report the simulation technique, the fabrication processes used and characterization of planar resonators having two different geometries. The fabricated resonators are used to record CW ESR spectrum of 2,2-diphenyl-1-picrylhydrazyl (DPPH) at room temperature and low temperature (77 K) in a home built setup.
\section{Simulation of the planar resonators}
\label{sim}
\begin{figure}[h!]
\centering
\includegraphics[scale=0.4]{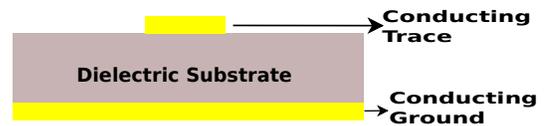}
\caption{Cross-sectional view of a microstrip line}
\label{fig:1}       
\end{figure}

\begin{figure}[!h]
\centering
\includegraphics[scale=0.5]{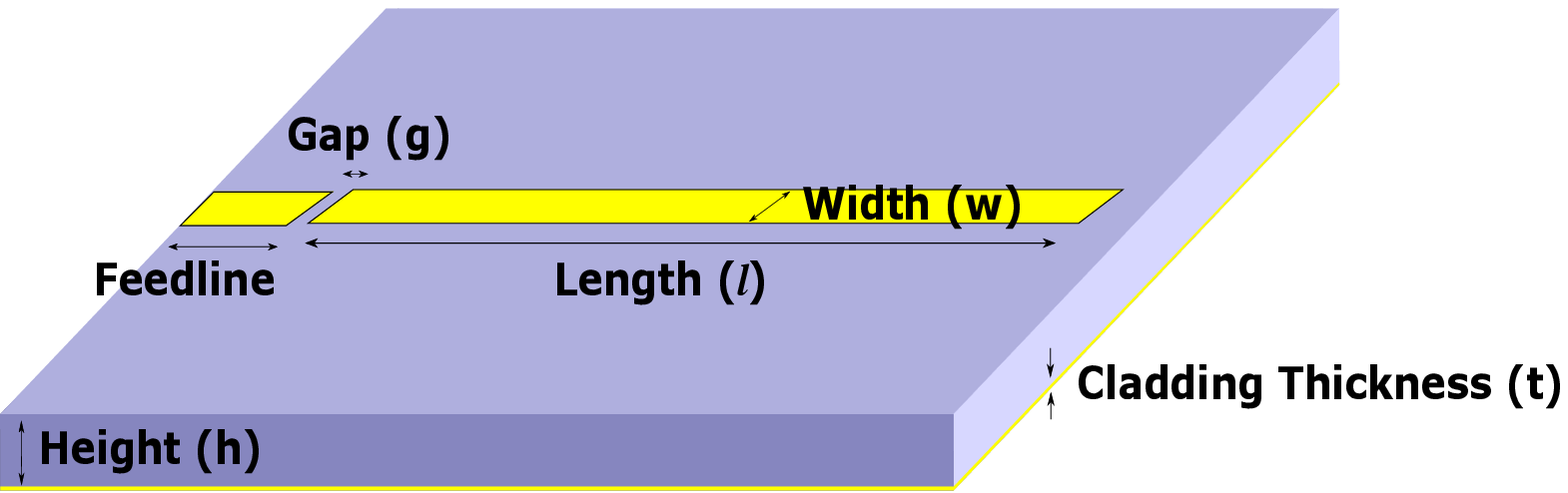}\\
\vspace{1 cm}
\includegraphics[scale=0.5]{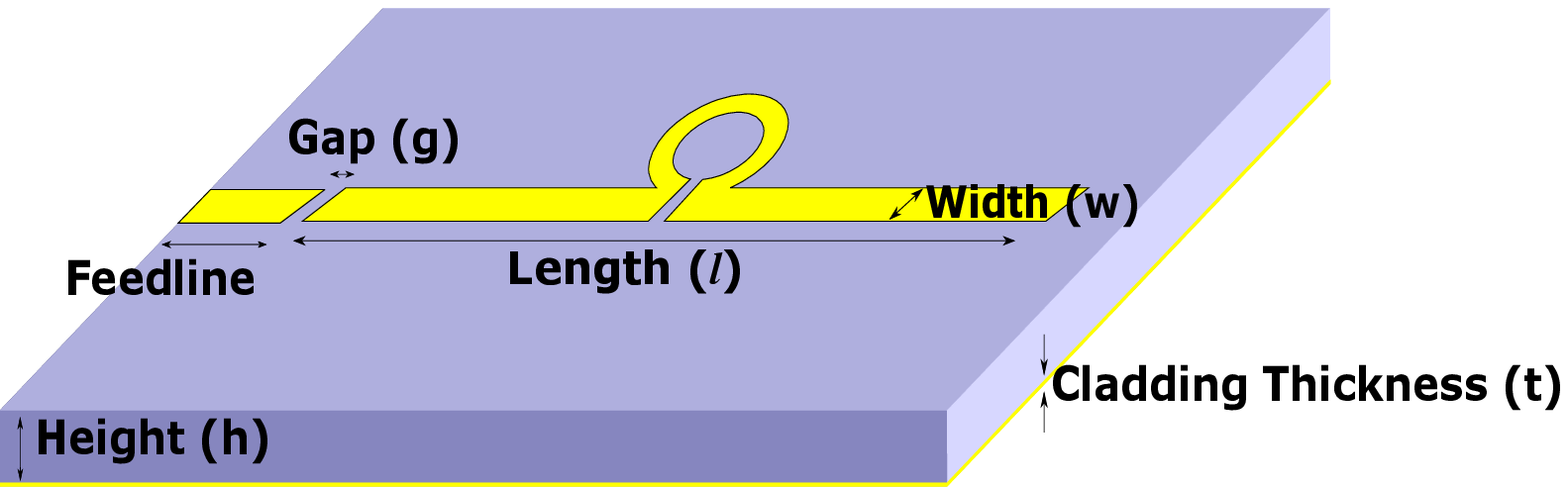}\\
\caption{Planar resonators - Straight Ribbon \& Omega}
\label{fig:2}       
\end{figure}

\begin{figure}
\centering
\includegraphics[width=0.6\columnwidth, height=0.6\columnwidth, trim=395 35 405 40,clip,scale=0.3]{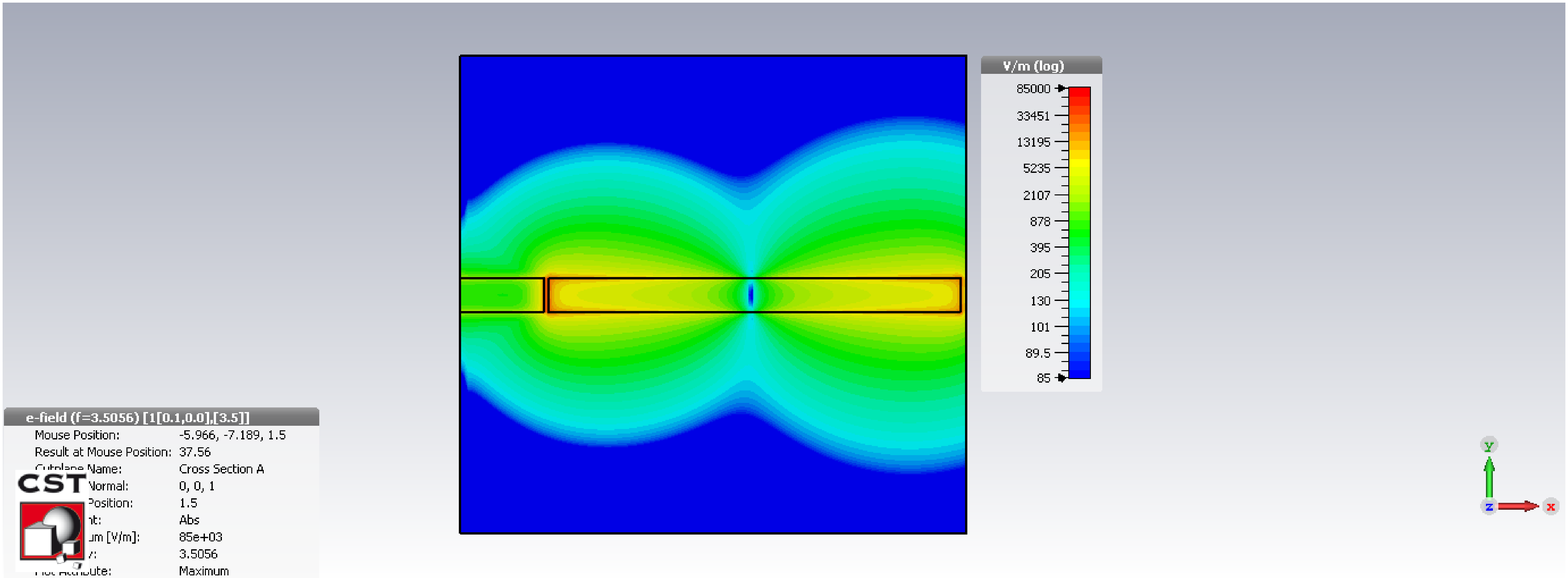}
\vspace{0.25 cm}
\includegraphics[width=0.6\columnwidth, height=0.6\columnwidth, trim=395 35 405 40,clip,scale=0.3]{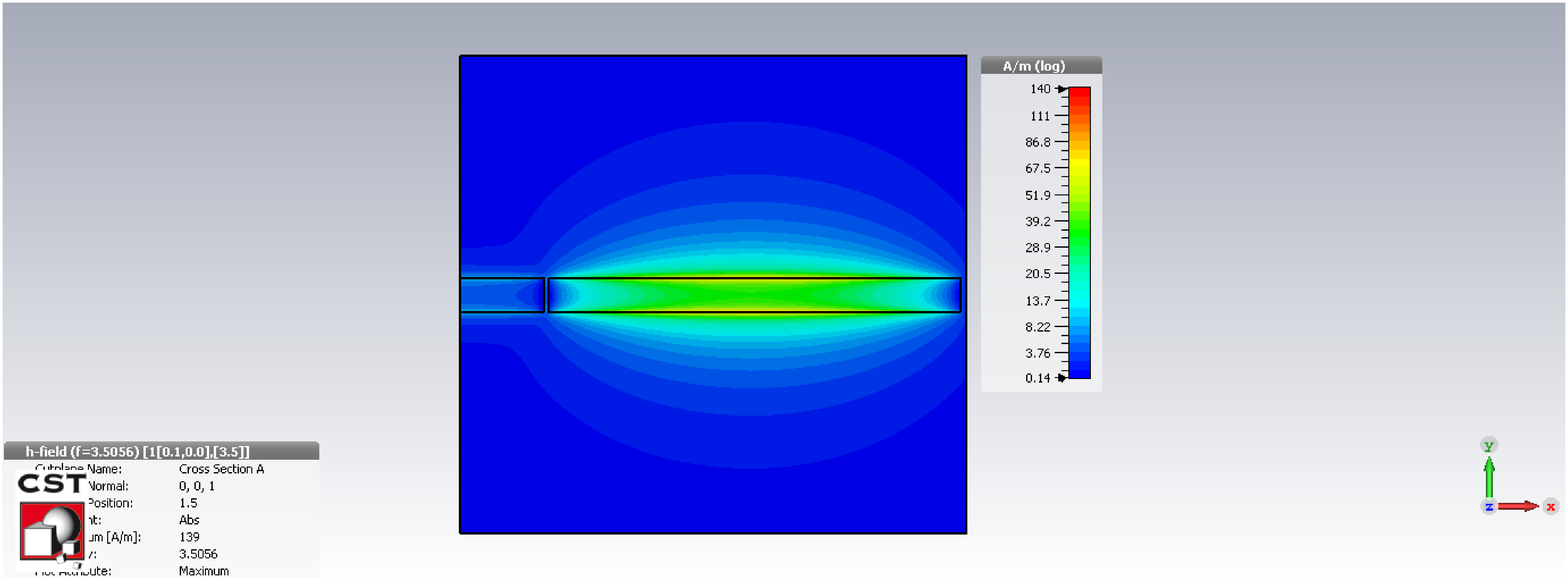}
\caption{Ribbon Resonator - Electric (top) and Magnetic (bottom) fields. [Input power = 10 dBm]}
\label{fig:3}       
\end{figure}

\begin{figure}
\centering
\includegraphics[width=0.6\columnwidth, height=0.6\columnwidth, trim=445 35 445 30,clip,scale=0.38]{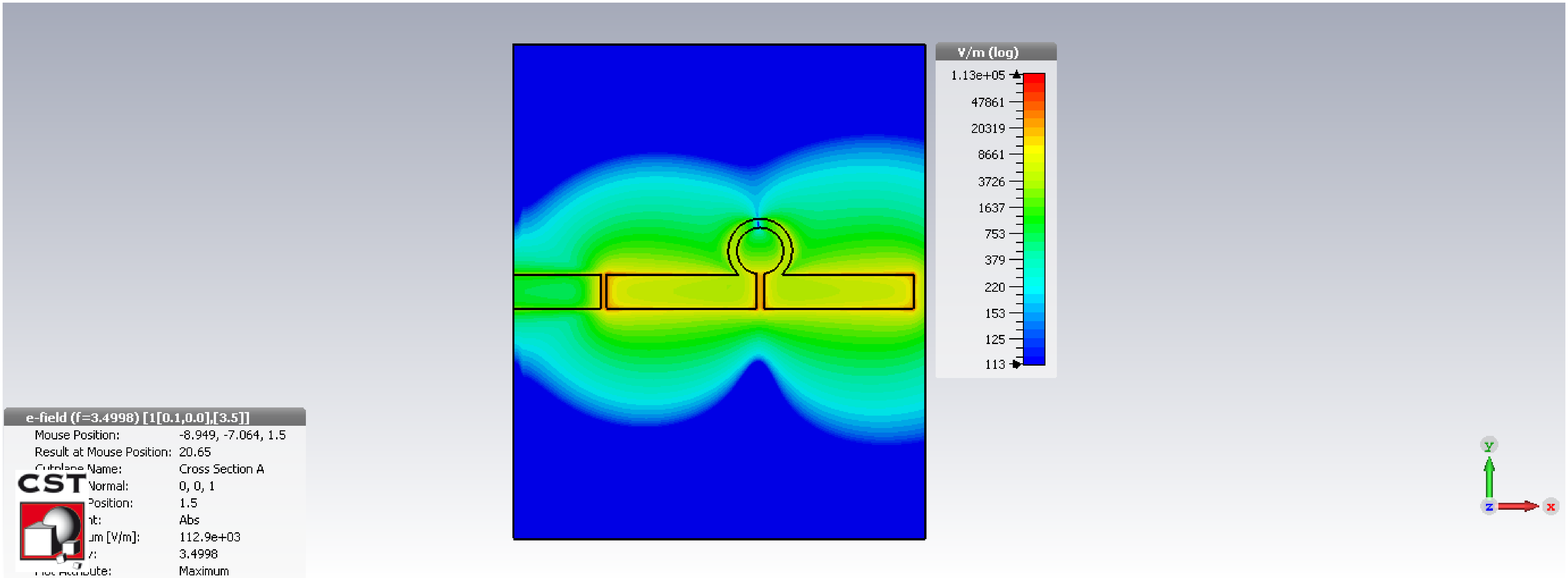}
\vspace{0.25 cm}
\includegraphics[width=0.6\columnwidth, height=0.6\columnwidth, trim=445 35 445 34,clip,scale=0.38]{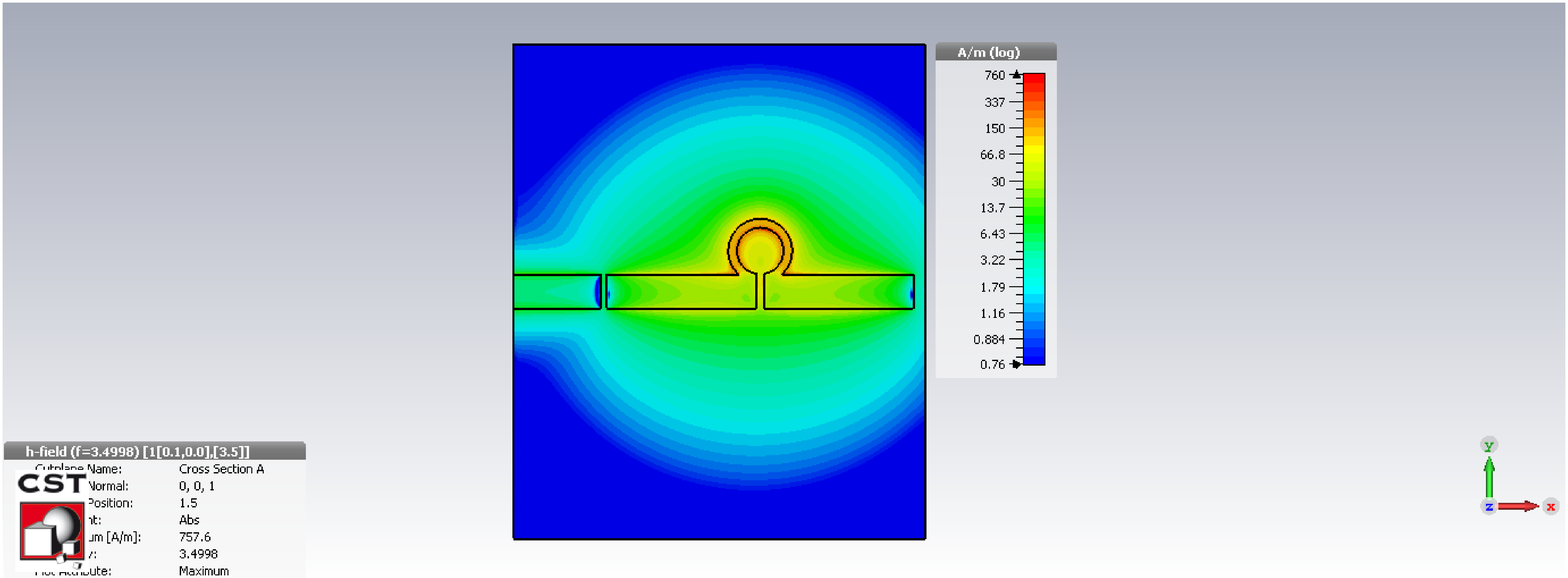}
\caption{Omega Resonator - Electric (top) and Magnetic (bottom) fields. [Input power = 10 dBm]}
\label{fig:4}       
\end{figure}

A microstrip line is a variant of the planar transmission line. It consists of three layers on top of one another - a bottom conducting ground plane, a middle dielectric substrate, and the conducting trace which embodies the design on top of the dielectric \cite{RefJ3}. The geometry of a microstrip line is indicated in figure \ref{fig:1}.
\\
The top conductor trace width $\mathbf{w}$ and the effective dielectric constant $\mathbf{\epsilon_{eff}}$ are analytically calculated by using the expressions given in  \cite{RefB3}. We have used AD1000 (Rogers Corporation), a commercial PTFE/woven glass based laminate with copper cladding on both sides of the substrate for design simulation and fabrication purpose. The dielectric is 1.5 mm thick (\textbf{h}) and has a relative permittivity ($\mathbf{\epsilon_{r}}$) of 10.7. A 17.5-micron thick copper cladding is present on either side of the substrate. One of them will act as the ground plane, and the resonator design will be etched on the other. It has a loss tangent (tan $\delta$)  of 0.0023. \\
We have simulated designs of planar half-wavelength resonators having two different geometries - straight ribbon \cite{RefJ4} and omega-shaped \cite{RefJ5}.A microstrip feed line which is incorporated on the substrate and separated from the planar resonator by a gap \textbf{g} couples the microwave to the resonators. The resonators were designed to resonate at 3.5 GHz (\textbf{f}\textsubscript{0}). The length of the resonators \textit{\textbf{l}} is analytically given by $\text{\textit{\textbf{l}}}+\Delta\text{\textit{\textbf{l}}} =\frac{\mathbf{\lambda_g}}{2}$ where $\mathbf{\lambda_g}$ is the guided wavelength and $\Delta\text{\textit{\textbf{l}}}$ is the correction term due to fringing electric fields and coupling gap. The geometries are indicated in the figure \ref{fig:2}.
\\
The electromagnetic simulations of the resonators are carried out in CST Microwave Studio software using time domain solver with open boundary condition.  The analytically obtained $\mathbf{w}$ and $\text{\textit{\textbf{l}}}$ are used as first approximation parameters in the simulation to set up the geometry of the resonators. The length \textit{\textbf{l}} was optimised to achieve a resonant frequency of the resonator close to $\mathbf{f_0}$. The coupling gap \textbf{g} was adjusted to create slight over coupling by keeping in mind the future use of the resonators in a pulsed ESR setup. The width \textbf{w} was tuned to adjust the characteristic impedance of the planar resonators to 50 $\Omega$ transmission line impedance. The design parameters obtained for both kind of resonators are listed in tables \ref{tab:1} \& \ref{tab:2}.
\begin{table}[!h]
\caption{Design Parameters - Straight Ribbon Resonator}
\label{tab:1}       
\centering
\begin{tabular}{ll}
\hline\noalign{\smallskip}
Parameter & Simulated  Result  \\
\hline\noalign{\smallskip}
Length of resonator (\textit{\textbf{l}}) & 14.67 mm  \\
Coupling Gap (\textbf{g}) & 0.15 mm \\
Width (\textbf{w}) & 1.18 mm \\
Feedline length & 3 mm \\
\noalign{\smallskip}\hline
\end{tabular}
\end{table}
\begin{table}[!h]
\caption{Design Parameters - Omega Resonator}
\label{tab:2}       
\centering
\begin{tabular}{ll}
\hline\noalign{\smallskip}
Parameter & Simulated  Result  \\
\hline\noalign{\smallskip}
Length of resonator (\textit{\textbf{l}}) & 14.26 mm  \\
Coupling Gap (\textbf{g}) & 0.2 mm \\
Width (\textbf{w}) & 1.18 mm \\
Omega gap & 0.25 mm \\
Omega inner radius & 0.8 mm\\
Omega width & 0.3 mm\\
Feedline length & 3 mm \\
\noalign{\smallskip}\hline
\end{tabular}
\end{table}\\
From the simulation we have obtained distribution of microwave electric field and magnetic field ($\mathbf{B_1}$) for both the resonators as shown in figures \ref{fig:3} \& \ref{fig:4}. It is seen from the simulation results that the microwave magnetic field strength for an omega resonator is almost one order magnitude higher than a ribbon resonator.
\section{Fabrication of the planar resonators}
\label{fab}
\begin{figure}[!h]
\centering
\includegraphics[scale=3.5]{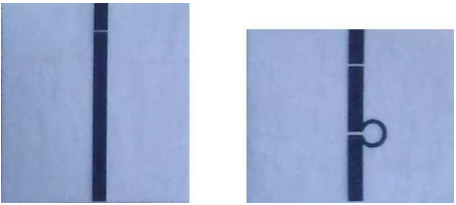}
\caption{Printed photomasks used in resonator fabrication}
\label{mask}       
\end{figure}

\begin{figure}[!h]
\centering
\includegraphics[scale=0.2]{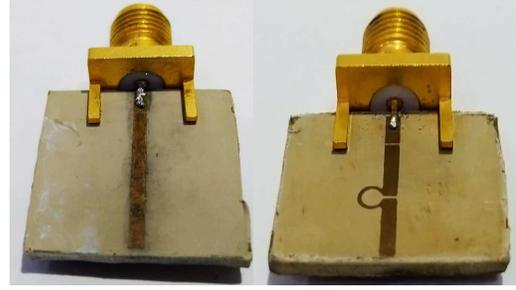}
\caption{Ribbon and Omega resonators after fabrication}
\label{fig:5}       
\end{figure}

The resonators are fabricated by utilizing a rapid prototyping method based on optical lithography \cite{RefJ6}-\cite{Rapid}. The simulated designs are exported from CST Microwave Studio  software to the computer-aided designing software Autodesk AutoCAD, in which electronic copy of photomasks is prepared. The designs are then printed on tracing paper (90 gsm thick) in a laser printer with a 1200 DPI resolution. The chosen resolution of the printer was able to resolve the smallest feature of the design.The printed photomasks are shown in figure \ref{mask}.
\\
The microwave laminate is cut into small rectangular pieces for making the planar resonators. The rectangular laminate pieces are cleaned with concentrated hydrochloric acid and acetone/isopropanol solution to remove the copper oxide layer and organic contamination respectively. Then their top surfaces are spin-coated with photoresist (S 1813) with the spin coater operated at 4000 rpm for 1 minute. The photoresist coated substrates are exposed to the ultraviolet (UV) light (365 nm, 18W) in a UV box in the presence of the printed photomasks for 5 minutes. The pattern is then developed with the developer (MF CD-26). Etching of the top surface is done with ferric chloride solution while the bottom copper surface is kept protected. End launch printed circuit board SMA connectors are then soldered to the feed line. The fabricated resonators are shown in figure \ref{fig:5}.

\section{Characterization of the planar resonators}
\begin{figure}
\centering
\includegraphics[width=0.75\columnwidth, height=0.75\columnwidth]{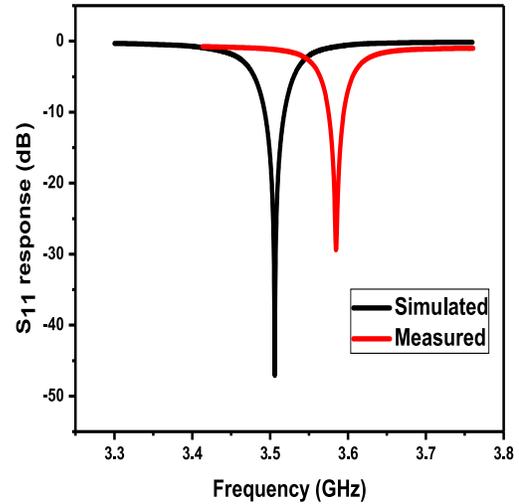}
\caption{Simulated (black) and measured (red) reflection co-efficient response of the ribbon resonator.}
\label{fig:first}
\end{figure}

\begin{figure}
\centering
\includegraphics[width=0.75\columnwidth, height=0.75\columnwidth]{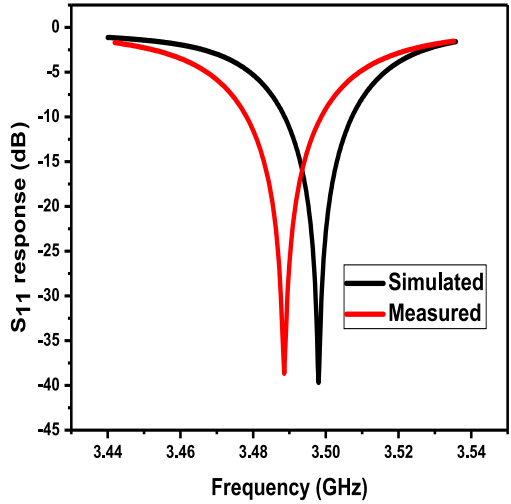}
\caption{Simulated (black) and measured (red) reflection co-efficient response of the omega resonator.}
\label{fig:second}
\end{figure}

\begin{figure}[!h]
\centering
\includegraphics[width=\columnwidth, height=0.85\columnwidth, trim=60 0 90 0,clip]{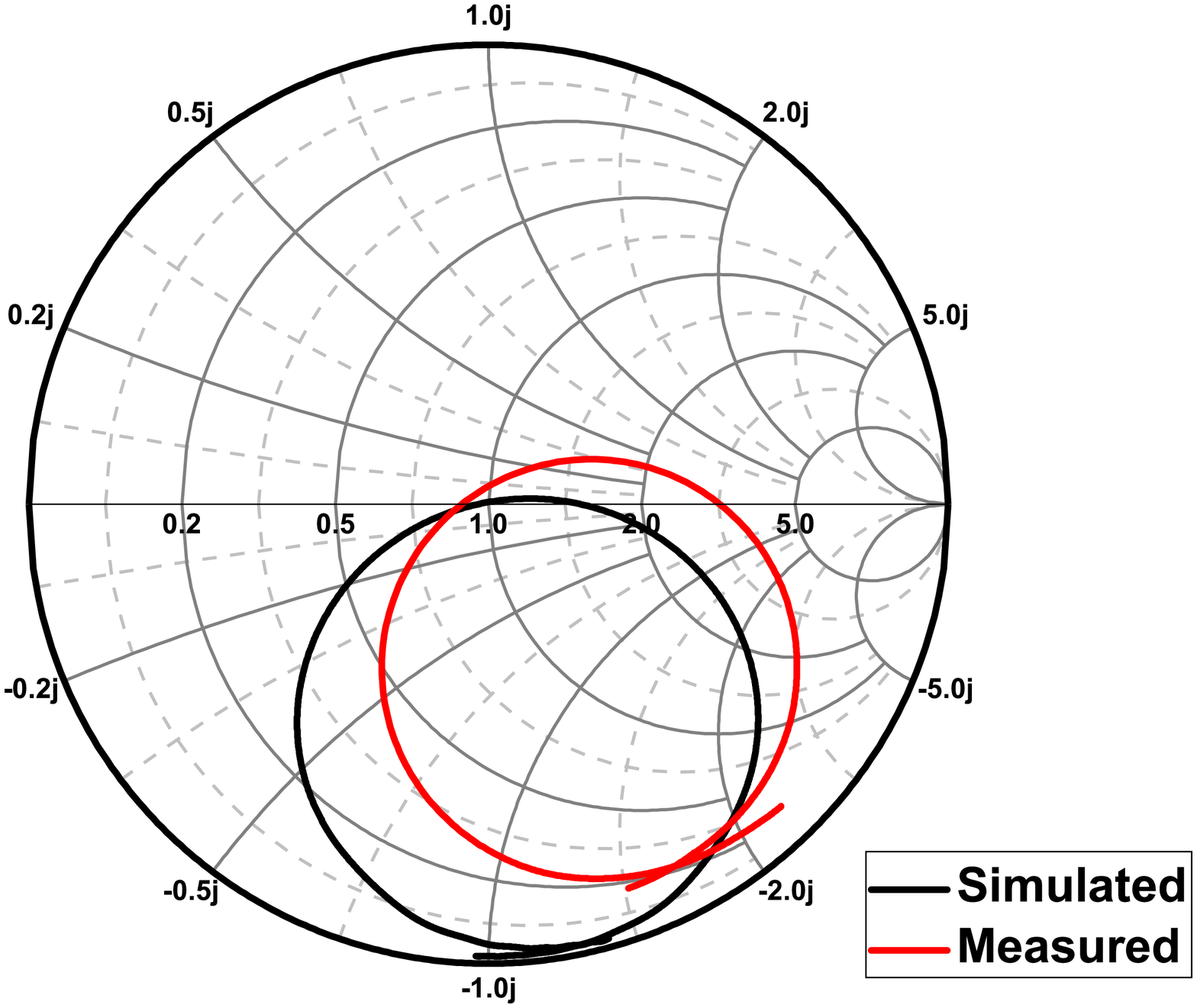}
\caption{Simulated (black) and measured (red) frequency sweep on the Smith chart of the ribbon resonator.}
\label{fig:third}
\end{figure}

\begin{figure}[!h]
\centering
\includegraphics[width=\columnwidth, height=0.85\columnwidth, scale=0.25,trim=60 0 85 0,clip]{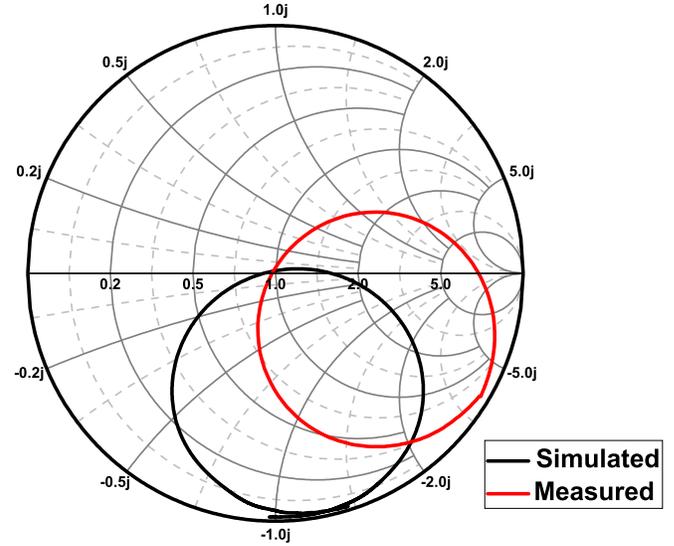}
\caption{Simulated (black) and measured (red) frequency sweep on the Smith chart of the omega resonator.}
\label{fig:fourth}
\end{figure}

Examination under high-resolution optical microscope (Olympus BX41M LED) shows good agreement between the fabricated resonators and simulated designs. The reflection co-efficient response in dBm for both the resonators are recorded using a Vector Network Analyzer (VNA) (ZVA 24, Rohde \& Schwarz) and compared to the simulated response curves as indicated in figures \ref{fig:first} and \ref{fig:second}. The resonant frequency response circle in the Smith chart shows that the resonators are slightly over-coupled \cite{RefB4} as shown in figures \ref{fig:third} and \ref{fig:fourth}.
\\
The resonant frequencies of the resonators are 3.58 GHz and 3.48 GHz for ribbon and omega resonators respectively. The loaded Q factor of the ribbon resonator was measured to be 67, and that of the omega resonator was 77 \cite{me}.
\section{Continuous Wave ESR - Instrumentation, Results and Discussion}
\begin{figure}
\centering
\includegraphics[scale=2.3]{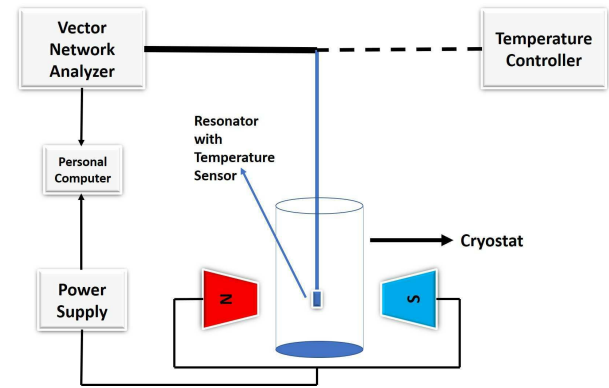}
\caption{Schematic of the home built ESR spectrometer}
\label{Schematic}       
\end{figure}

\begin{figure}
\centering
\includegraphics[scale=0.27,trim=40 0 95 0,clip]{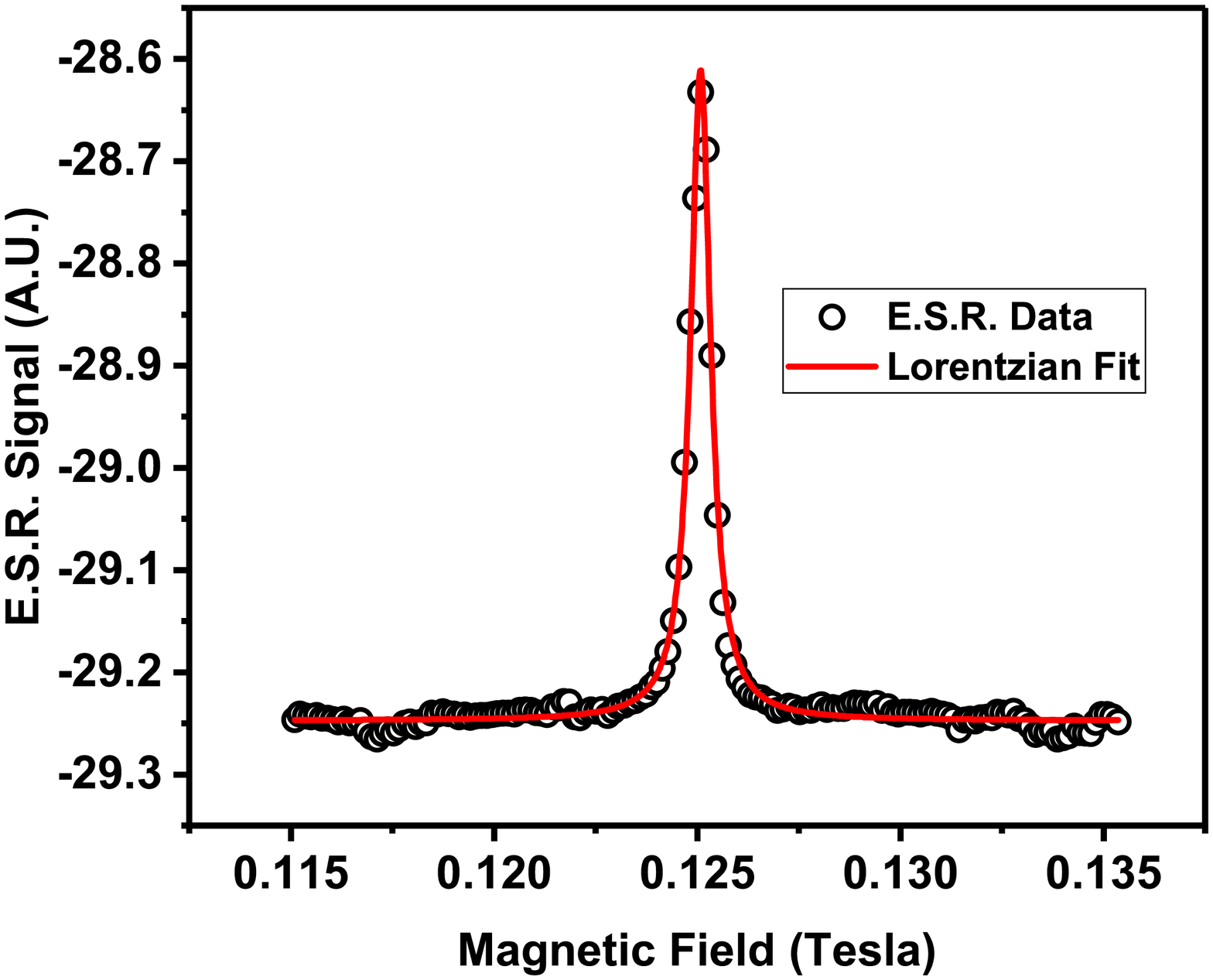}
\caption{Room temperature ESR absorption spectrum of DPPH captured using ribbon resonator. The solid red curve depicts the Lorentzian fit. (Input Power = 18 dBm)}
\label{esr_half}       
\end{figure}

\begin{figure}
\centering
\includegraphics[scale=0.27,trim=20 0 85 0,clip]{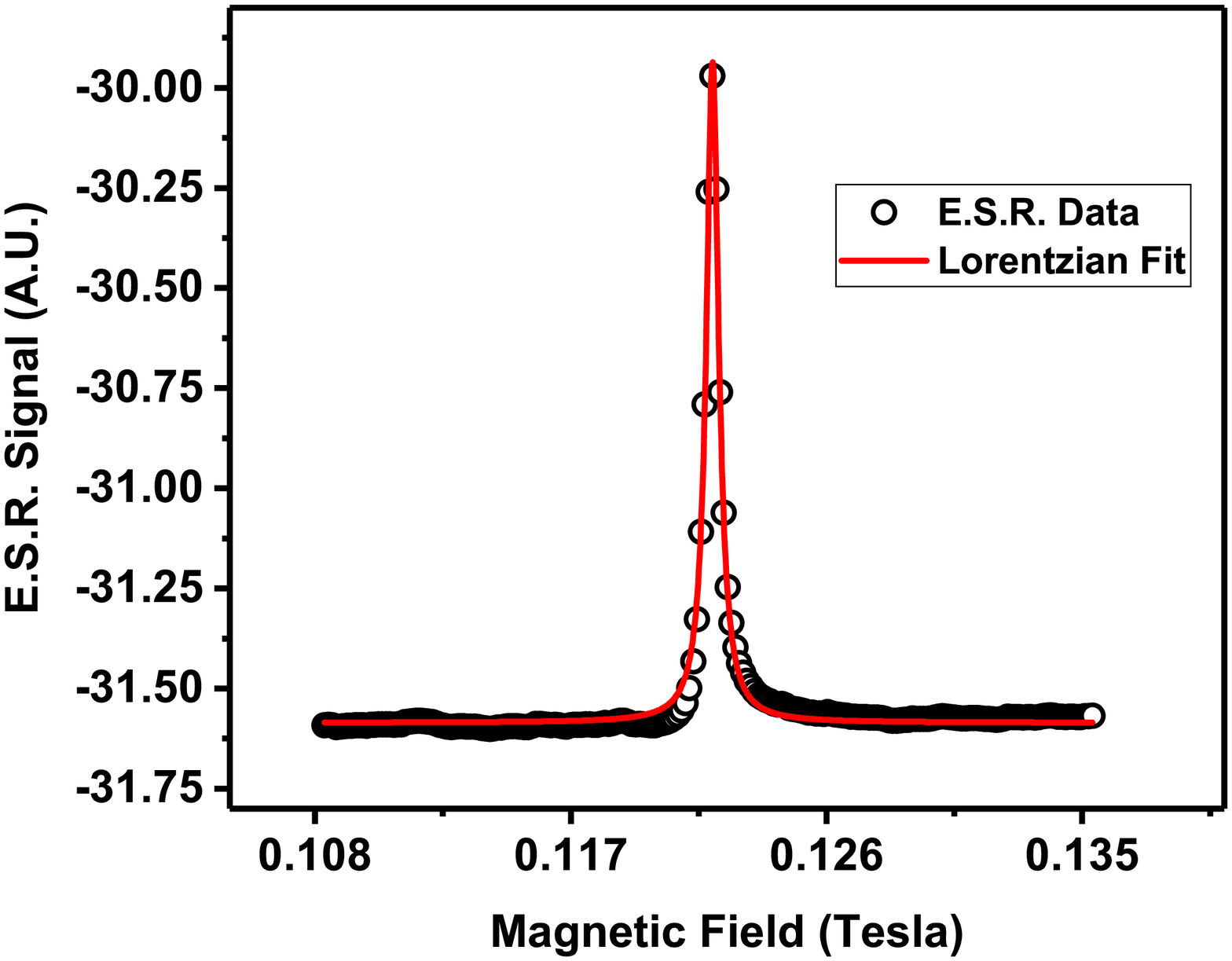}
\caption{Room temperature ESR absorption spectrum of DPPH captured using omega resonator. The solid red curve depicts the Lorentzian fit. (Input Power = 18 dBm)}
\label{esr_omega}       
\end{figure}

\begin{figure}[!h]
\centering
\includegraphics[scale=0.27,trim=60 0 80 0,clip]{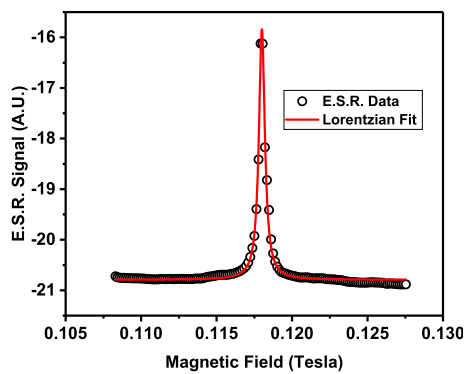}
\caption{Low temperature (77 K) ESR absorption spectrum of DPPH captured using omega resonator. The solid red curve depicts the Lorentzian fit. (Input Power = 18 dBm)}
\label{low_temp}       
\end{figure}

We have performed continuous wave ESR measurement on DPPH in a home built setup as indicated in figure \ref{Schematic}. The spectrometer is based on single port measurement geometry. The setup consists of a VNA which acts as a source and detector of microwave signals and an electromagnet (GMW  3473-70) having a programmable power supply (Sorensen SGA60X83D) for generating the Zeeman field (\textbf{B\textsubscript{0}}). The field of the electromagnet is calibrated against the power supply's current by using a gaussmeter (DTM-151, GMW Associates). Semi-rigid coaxial transmission lines (RMSC-033/50-SS-SS, Response Microwave Inc.) are used to connect the resonator to the VNA. The sample wrapped in Teflon tape is placed at the center of the resonator in case of ribbon resonator or inside the loop in case of the omega resonator during room temperature measurements. These are the positions of \textbf{B\textsubscript{1}} antinode. \\
Low-temperature measurements are carried out in a custom made liquid nitrogen cryostat. The temperature controller (Cryocon 24C) senses the temperature through a platinum RTD temperature sensor (Cryocon XP-100). The omega resonator was used while doing low-temperature measurements since it has higher  \textbf{B\textsubscript{1}} flux density as compared to the ribbon resonator. The resonator and the temperature sensor are mounted on a custom designed low temperature insert while doing measurements at 77 K.\\
The insert consists of a semirigid coaxial cable that is affixed through a rubber o-ring and Wilson seal at the top of the insert, for holding the coaxial cable firmly in place and for maintaining a vacuum. The SMA connectors to the top and bottom of the semi-rigid cable were soldered after the cable was passed through the o-ring and the Wilson seal. It has the option of sliding the semirigid cable to attach the sample in the middle of the electromagnet. The whole cable assembly was affixed inside seamless stainless steel (SS) tube, which was again passed through a larger Wilson seal at the top to slide it inside the cryostat. The cable assembly was firmly held at the bottom of the seamless SS tube through screws. OSM calibration was performed before the resonators were attached to the bottom of the semi-rigid cable of the insert.The sample is held on the \textbf{B\textsubscript{1}} anti-nodal position of the resonator with the help of a low-temperature Kapton tape. \\ 
The resonators are positioned inside the electromagnet in such a manner so that \textbf{B\textsubscript{1}} is perpendicular to \textbf{B\textsubscript{0}}.The device interfacing program was written in Python. The room temperature spectra recorded by using both the resonators are shown in figures \ref{esr_half} \& \ref{esr_omega}. The low- temperature spectra recorded by using omega resonator is shown in figure \ref{low_temp}. 
\\
The calculated g-factor of DPPH was 2.04 in room temperature and 2.02 at low temperature. The measured g-factor values are slightly higher than the literature value \cite{RefJ7}. This can be attributed to the limitation of the power supply's accuracy and tolerance of the fabrication procedure. The recorded ESR spectra showed a good fit to a standard Lorentzian function. The signal to noise ratio (SNR)\cite{RefJ8} of the spectrometer is found to be 80 at room temperature if ribbon resonator is used. On using the omega resonator SNR improved to 204 at room temperature and to 481 at 77 K.
\section{Conclusion}
The planar resonators with two different geometries have been successfully fabricated by using photomask printed in normal laser printer at a high resolution. ESR signal of a standard sample has been well captured by using our home built setup at both room temperature and low temperature with a good SNR. The resonators will be used in future as a component of pulsed ESR spectrometer.The simple and low cost fabrication procedure with a low turnover time (few hours) of each resonator yielding a significant signal to noise ratio makes them very useful and the research done in this work is of significant importance. 
\section*{Authors' Contribution}
\textsuperscript{a}Both authors have contributed equally towards this work.
\section*{Acknowledgements}
Authors acknowledge Ministry of Human Resource Development (MHRD), Government of India \& Science and Engineering Research Board (SERB) (grant no. - EMR/2016/007950) for funding this work. S. R. acknowledges Council of Scientific \& Industrial Research (CSIR), India for research fellowship. J.S acknowledges the University Grants Commission (UGC), India  for research fellowship. The authors acknowledge the valuable insights given by Dr. Partha Pratim Sarkar of the Department of Engineering and Technological Studies (DETS), University of Kalyani. The authors thank Prof. Bhaskar Gupta, Department of Electronics \& Telecommunication Engineering , Jadavpur University for providing simulation facilities. The authors are grateful to Roger Corporation, USA for free samples of the microwave laminate.


\end{document}